\documentclass[a4paper,11pt]{article}
\usepackage{amsmath}
\usepackage{graphicx,psfrag,epsf}
\usepackage{enumerate}
\usepackage{url} 
\usepackage{amssymb}
\usepackage{algpseudocode}
\usepackage{algorithmicx}
\usepackage{algorithm2e}
\newtheorem{lemma}{Lemma}

\usepackage{color}
\usepackage[export]{adjustbox}
\newcommand*{\mathcolor}{}
\def\mathcolor#1#{\mathcoloraux{#1}}
\newcommand*{\mathcoloraux}[3]{%
  \protect\leavevmode
  \begingroup
    \color#1{#2}#3%
  \endgroup
}
\usepackage[usenames,dvipsnames,svgnames,table]{xcolor}
\usepackage[colorlinks]{hyperref}
\hypersetup{
  colorlinks=true,
  linkcolor=red,
  citecolor=red
}
\AtBeginDocument{%
  \hypersetup{%
    linkcolor=red,%
    citecolor=blue,%
  }%
}%

\def\bSig\mathbf{\Sigma}

\usepackage[left=3cm,right=3cm,top=2cm,bottom=2cm]{geometry}

\begin{document}
\date{}

\title{\bf Extension to mixed models of the Supervised Component-based Generalised Linear Regression}
\author{
\textbf{Jocelyn Chauvet}$^{1}$ \and
\textbf{Catherine Trottier}$^{1,2}$ \and
\textbf{Xavier Bry}$^{1}$ \and 
\textbf{Fr\'ed\'eric Mortier}$^{3}$ \\ \\
$^{1}$ Institut Montpelli\'erain Alexander Grothendieck, \\ CNRS, Univ. Montpellier, France. \\ 
\small{\texttt{jocelyn.chauvet@umontpellier.fr}, 
\texttt{xavier.bry@umontpellier.fr}} \\
$^{2}$ Univ. Paul-Val\'ery Montpellier 3, \\ F34000, Montpellier, France. \\
\small{\texttt{catherine.trottier@univ-montp3.fr}}   \\
$^{3}$ Cirad, UR BSEF, Campus International de Baillarguet \\ TA C-105/D - 34398 Montpellier, \\ \small{\texttt{frederic.mortier@cirad.fr}} 
}
\maketitle


\begin{abstract}
We address the component-based regularisation of a multivariate Generalized Linear Mixed Model (GLMM).  
A set of random responses $Y$ is modelled by a GLMM, using a set $X$ of explanatory variables, a set $T$ of additional covariates, and random effects used to introduce the dependence between statistical units.
Variables in $X$ are assumed many and redundant, so that regression demands regularisation. 
By contrast, variables in $T$ are assumed few and selected so as to require no regularisation.
Regularisation is performed building an appropriate number of orthogonal components that both contribute to model $Y$ and capture relevant structural information in $X$.
To estimate the model, we propose to maximise a criterion specific to the Supervised Component-based Generalised Linear Regression (SCGLR) within an adaptation of Schall's algorithm.
This extension of SCGLR is tested on both simulated and real data, and compared to  Ridge- and Lasso-based regularisations.
\end{abstract}


\noindent%
\vfill

\newpage
\section{Data, Model and Problem}
\paragraph{}
A set of $q$ random responses 
$Y = \left\lbrace y^1, \ldots, y^q \right\rbrace$ 
is assumed explained by two different sets of covariates 
$X = \left\lbrace x^1, \ldots, x^p \right\rbrace$ and 
$T = \left\lbrace t^1, \ldots, t^r \right\rbrace$, and a random effect 
$\xi = \left\lbrace \xi^1, \ldots, \xi^q \right\rbrace$.
Explanatory variables in $X$ are assumed many and redundant while additional covariates in $T$ are assumed selected so as to preclude redundancy.
Explanatory variables in $T$ are thus kept as such in the model. 
By contrast, $X$ may contain several unknow structurally relevant dimensions $K<p$ important to model and predict $Y$, how many we do not know.
$X$ is thus to be searched for an appropriate number of orthogonal components that both capture relevant structural information in $X$ and contribute to model $Y$.

\vspace{3mm}
Each $y^k$ is modelled through a Generalised Linear Mixed Model (GLMM) 
\cite{McCullochSearle:2001} 
assuming conditional distributions from the exponential family.
More specifically in this work, the $n$ statistical units are not considered independent but partitioned into $N$ groups. 
The random effects included in the GLMM aim at modelling the dependence of units within each group.

\vspace{3mm}
Over the last decades, component-based regularisation methods for Generalised Linear Models (GLM) have been developped.
In the univariate framework, i.e. when $Y = \left\lbrace y \right\rbrace$, 
Bastien et al. \cite{Batienetal:2004} 
proposed an extension to GLM of the classical Partial Least Square (PLS) regression, combining generalised linear regressions of the dependent variable on each of the regressors considered separately.
However, doing so, this method does not take into account the variance structure of the overall model when building a component.
Still in the univariate framework, 
Marx \cite{Marx:1996} 
proposed a more appropriate Iteratively Reweighted Partial Least Squares (IRPLS) estimation that builds PLS components using the weighting matrix derived from the GLM.
More recently, 
Bry et al. \cite{Bryetal:2013} 
extended the work by 
Marx \cite{Marx:1996} 
to the multivariate framework with a technique named Supervised Component-based Generalised Linear Regression (SCGLR). The basic principle of SCGLR is to build optimal components common to all the dependent variables. To achieve it, SCGLR introduces a new criterion which is maximised at each step of the Fisher Scoring Algorithm (FSA).

\vspace{3mm}
Besides, regularisation methods have already been developped for GLMM, in which the random effects allow to model complex dependence structure.
Eliot et al. \cite{Eliotetal:2011} proposed to extend the classical ridge regression to Linear Mixed Models (LMM). The Expectation-Maximisation algorithm they suggest includes a new step to find the best shrinkage parameter - in the Generalised Cross-Validation (GCV) sense - at each iteration. 
More recently, 
Groll and Tutz \cite{GrollandTutz:2014} proposed an $L_1$-penalised algorithm for fitting a high-dimensional GLMM, using Laplace approximation and efficient coordinate gradient descent.

\vspace{3mm}
Instead of using a penalty on the norm of the coefficient vector, we propose to base the regularisation of the GLMM estimation on SCGLR-type components.



\section{Reminder on SCGLR with additional covariates}
\paragraph{}
In this section, we consider the simplified situation where each $y^k$ is modelled through a GLM (without random effect) and only one component is calculated ($K=1$). Moreover, let us use the following notations:

$\Pi_E^M$ : orthogonal projector on space $E$, with respect to some metric $M$.

$\langle X \rangle$ : space spanned by the column-vectors of $X$.

$M'$ : transpose of any matrix (or vector) $M$.

\paragraph{}
The first conceptual basis of SCGLR consists in searching for an optimal component $f=Xu$ common to all the $y$'s.
Therefore, SCGLR adapts the classical FSA to predictors having colinear $X$-parts. To be precise, for each $k \in \left\lbrace 1, \ldots, q\right\rbrace$, the linear predictor writes:
\begin{equation*} 
\eta^k = (Xu)\gamma_k + T\delta_k 
\end{equation*}
where $\gamma_k$ and $\delta_k$ are the parameters associated with component $f=Xu$ and covariates $T$ respectively.
For identification, we impose $u'Au=1$, where $A$ may be any symmetric definite positive matrix.
Assuming that both the $y$'s and the $n$ statistical units are independent, the likelihood function $L$ can be written:
\begin{equation*}
L(y|\eta) = \prod_{i=1}^n \prod_{k=1}^q L_k (y_i^k | \eta_i^k)
\end{equation*}
where $L_k$ is the likelihood function relative to $y^k$.
Owing to the product $\gamma_k u$, the ``linearised model'' (LM) on each step of the associated FSA for the GLM estimation is not indeed linear: an alternated least squares step was designed.
Denoting $z^k$ the classical working variables on each FSA's step and $W_k^{-1}$ their variance-covariance matrix, the least squares on the LM consists in the following optimisation (see \cite{Bryetal:2013}): 
\begin{equation*}
\underset{u'Au=1}{\min} \; 
\sum_{k=1}^q  \left\lVert z^k - \Pi_{\langle Xu,T \rangle}^{W_k} z^k \right\rVert_{W_k}^2 
\quad  \Longleftrightarrow  \quad
\underset{u'Au=1}{\max}  \;
\sum_{k=1}^q  \left\lVert \Pi_{\langle Xu,T \rangle}^{W_k} z^k \right\rVert_{W_k}^2
\end{equation*}
which is also equivalent to : 
\begin{equation}
\underset{u'Au=1}{\max} \; \psi_T(u), \quad
\text{with} \quad \psi_T(u) = \sum_{k=1}^q \left\lVert z^k \right\rVert_{W_k}^2 
\cos_{W_k}^2 \left( z^k, \langle Xu, T \rangle \right)
\label{Jocelyn.Chauvet:GoF_criterion}
\end{equation}

\begin{figure}[ht]
\begin{center}
\includegraphics[width=\linewidth]{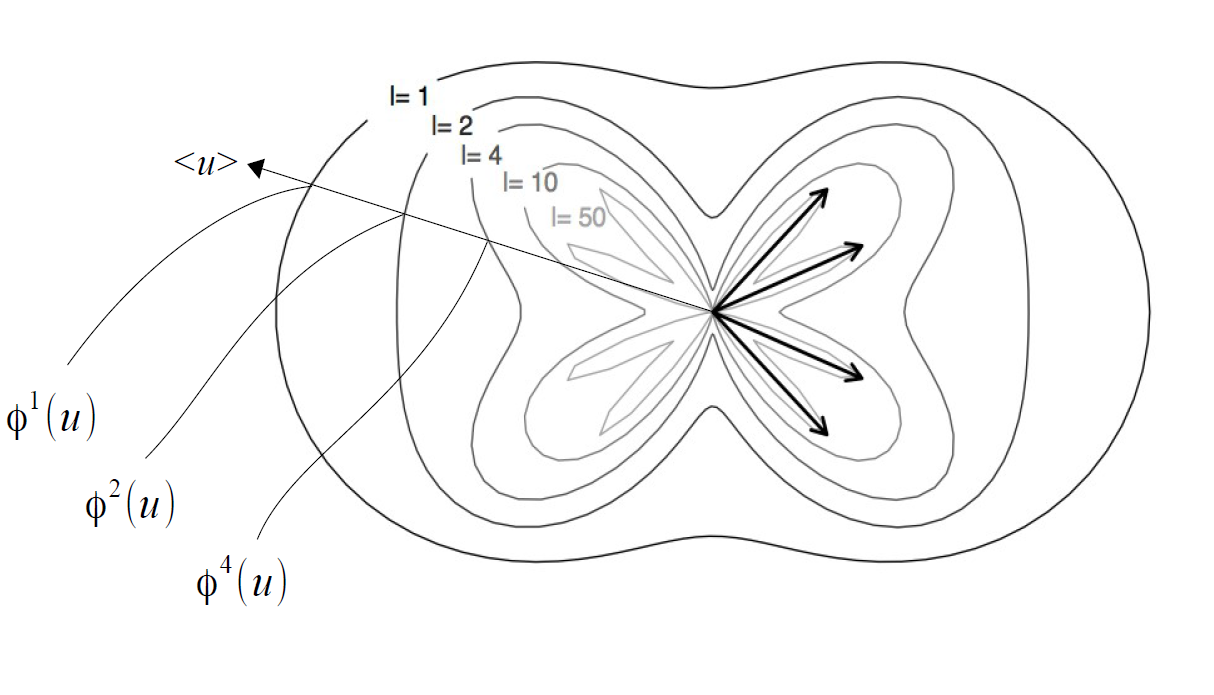}
\end{center} 
\vspace{-2em}
\caption{Polar representation of $\phi^l(u)$ according to the value of $l$, in the elementary case of four coplanar variables.}
\label{Jocelyn.Chauvet:locality_bundles}
\end{figure}

\paragraph{}
The second conceptual basis of SCGLR consists in introducing a closeness measure of the component $f=Xu$ to the strongest structures in $X$. Indeed, $\psi_T$ is a mere goodness-of-fit measure, and must be combined with a structural relevance measure to get regularisation. 
Consider a given weight-matrix $W$ - e.g. $W=\frac{1}{n} I_n$ - reflecting the a priori relative importance of units, the most structurally relevant component would be the solution of:
\begin{equation}
\underset{u'Au=1}{\max} \; \phi(u), \quad
\text{with} \quad
\phi(u) 
= \left(    {\overset{p}{\underset{j=1}{\sum}}}  \langle Xu | x^j \rangle_{W}^{2l}      
\right)^{\frac{1}{l}}
= \left(   {\overset{p}{\underset{j=1}{\sum}}} 
(u' X' W x^j {x^j}' W X u)^{l}      
\right)^{\frac{1}{l}} 
\label{Jocelyn.Chauvet:SR_criterion}
\end{equation}
Tuning parameter $l$ allows to draw components towards more (greater $l$) or less (smaller $l$) local variable bundles as depicted on \autoref{Jocelyn.Chauvet:locality_bundles}.

\paragraph{}
To sum things up, 
$s$ being a parameter that tunes the importance of the structural relevance relative to the goodness-of-fit,
SCGLR attempts a trade-off between (\ref{Jocelyn.Chauvet:GoF_criterion}) and (\ref{Jocelyn.Chauvet:SR_criterion}), solving:
\begin{equation}
\underset{u'Au=1}{\max} \; \left[\phi(u)\right]^s \: \left[\psi_T(u)\right]^{1-s}
\label{Jocelyn.Chauvet:CompromiseCriterion}  
\end{equation}



\section{Adapting SCGLR to grouped data}

\paragraph{}
We propose to adapt SCGLR to grouped data, for which the independence assumption of the statistical units is no longer valid (e.g. longitudinal or spatial correlated data).
The whithin-group dependence is modelled by a random effect.
Consequently, each of the $y^k$ is ultimately modelled through a GLMM.
We call this adaptation ``Mixed-SCGLR".

\subsection{Single component model estimation}
\paragraph{}
We first present the method's principle. Then we give a Bayesian justification of the involved Henderson systems. Finally, we present our algorithm's steps.

\paragraph{Principle}
We maintain predictors colinear in their $X$-parts. 
Introducing a random effect in each predictor and still imposing $u'Au=1$, the linear predictors now write:
\begin{equation}
\forall k \in \left\lbrace 1, \ldots, q\right\rbrace, \;\;
\eta_{\xi}^k = (Xu)\gamma_k + T\delta_k + U\xi^k
\label{Jocelyn.Chauvet:random_predictor}
\end{equation}
The random group effect is assumed different accross responses, yielding $q$ random effects 
$\xi^1, \ldots, \xi^q$. These are assumed independent and normally distributed: 
$\mathcal{N}_N(0, D_k = \sigma_k^2 A_k)$, where
$N$ is the number of groups and $A_k$ a known matrix ($A_k = I_N$ in general).

\vspace{3mm}
Owing to the GLMM dependence structure, the FSA was adapted by Schall \cite{Schall:1991}. We, in turn, adapt Schall's algorithm to our component-based predictor (\ref{Jocelyn.Chauvet:random_predictor}), by introducing the following alternated procedure at each step:
\begin{itemize}
\item
Given $\gamma_k$, $\delta_k$, $\xi^k$ and $\sigma_k^2$, 
we build the component $f=Xu$ by solving a (\ref{Jocelyn.Chauvet:CompromiseCriterion})-type program, which attempts a compromise between goodness-of-fit and structural relevance criterions.
\item
Given $u$, 
$z_{\xi}^k$ being the classical working variables of the Schall's algorithm and 
$W_{\xi,k}^{-1}$ their conditional variance-covariance matrix,
parameters $\gamma_k$, $\delta_k$ and $\xi^k$ are estimated by solving the following Henderson system, which, subsequently, allows us to estimate 
$\sigma_k^2$ : 
\begin{equation}
\begin{pmatrix}
(Xu)'W_{\xi,k}(Xu) & (Xu)'W_{\xi,k}T & (Xu)'W_{\xi,k}U \smallskip \\
T'W_{\xi,k}(Xu)    & T'W_{\xi,k}T    & T'W_{\xi,k}U    \smallskip \\
U'W_{\xi,k}(Xu)    & U'W_{\xi,k}T    & U'W_{\xi,k}U + D_k^{-1}
\end{pmatrix}
\begin{pmatrix}
\gamma_k \smallskip \\ 
\delta_k \smallskip \\ 
\xi^k
\end{pmatrix} 
= 
\begin{pmatrix}
(Xu)'W_{\xi,k} \, z_{\xi}^k \smallskip \\
T'W_{\xi,k} \, z_{\xi}^k    \smallskip \\
U'W_{\xi,k} \, z_{\xi}^k
\end{pmatrix}
\label{Jocelyn.Chauvet:Henderson-system}
\end{equation}
We chose Henderson's method \cite{Henderson:1975} since it is quicker than EM, for instance.
\end{itemize}

\paragraph{A Bayesian justification of the Henderson systems}
The conditional distribution of the data, given the random effects, is supposed to belong to the exponential family, i.e.  for each $k \in \left\lbrace 1, \ldots, q \right\rbrace$, the conditional density of $Y_i^k \,|\, \xi^k$ may be written:
\begin{equation*}
{f_{}}_{Y_i^k | \xi^k} \left( y_i^k, \theta_i^k\right) = 
\exp \left\lbrace
\dfrac{y_i^k \theta_i^k - b_k\left(\theta_i^k\right)}{a_{k,i}(\phi_k)}  + c_k\left(y_i^k,\phi_k\right)
\right\rbrace
\end{equation*}

\noindent
\textit{Linearisation step} :
Denoting $G_k$ the link function of variable $y^k$, 
$g_k$ its first derivative
and $\mu_k$ the conditional expectation 
(i.e. $\mu^k := \mathbb{E}(Y^k \, | \, \xi^k)$),
the working variables are obtained as:
\begin{equation*}
z_{\xi}^k = \eta_{\xi}^k + e^k, \quad
\text{where} \quad e_i^k = (y_i^k - \mu_i^k) g_k(\mu_i^k)
\end{equation*}
Their conditional variance-covariance matrices are:
\begin{equation*}
W_{\xi,k}^{-1} := \mbox{Var} \left(z_{\xi}^k \,| \, \xi^k \right) = \mbox{Diag} \left(
\left[g_k\left(\mu_i^k \right)\right]^2 
\mbox{Var} \left(  Y_i^k \,| \, \xi^k \right)
\right)_{i=1, \ldots, n}
\end{equation*}

\noindent
\textit{Estimation step} : 
Our estimation step is based on the following lemma about normal hierarchy:

\medskip
\hrule
\begin{lemma}
Given
\begin{align*}
y|\theta &\sim \mathcal{N}(M\theta, R) \\
\theta &\sim \mathcal{N}(\alpha, \Omega)
\end{align*}
the posterior distribution is 
$\theta|y \sim \mathcal{N}(\hat{\theta}, C)$, 
where 
$C = \left( M'R^{-1}M + \Omega^{-1}\right)^{-1}$
and $\hat{\theta}$ satisfies:
\begin{equation}
C^{-1} \hat{\theta} = M'R^{-1}y + \Omega^{-1} \alpha. 
\label{Jocelyn.Chauvet:equation_lemma}
\end{equation}
\end{lemma}
\hrule
\bigskip
Given $u$, 
we just apply Schall's method \cite{Schall:1991} with our regularised linear predictors (\ref{Jocelyn.Chauvet:random_predictor}), which is equivalent to   consider the following modelling:
\begin{align}
z_{\xi}^k | \gamma_k, \delta_k, \xi^k &\sim 
\mathcal{N} \left( (Xu)\gamma_k + T\delta_k + U\xi^k, \; W_{\xi,k}^{-1}  \right) 
\label{Jocelyn.Chauvet:working_var_assuption} \\
\left( \gamma_k, \delta_k, \xi^k \right)  
&\sim \mathcal{N} \left( 
\begin{pmatrix}
\gamma_k^{(0)} \\ \delta_k^{(0)} \\ 0
\end{pmatrix}, \; 
\begin{pmatrix}
B_k^{\gamma} & 0 & 0 \\
0 & B_k^{\delta} & 0 \\
0 & 0 & D_k
\end{pmatrix} \nonumber
\right)
\end{align}
We suggest to choose a noninformative prior distribution for parameters $\gamma_k$ et $\delta_k$ (as inspired by Stiratelli et al. \cite{Stiratellietal:1984}) imposing 
${B_k^{\gamma}}^{-1} = {B_k^{\delta}}^{-1} = 0$.
Current estimates of parameters $\gamma_k, \delta_k$ and $\xi^k$ are thus obtained solving (\ref{Jocelyn.Chauvet:equation_lemma}), which is equivalent to the Henderson system (\ref{Jocelyn.Chauvet:Henderson-system}).

\vspace{3mm}
Finally, as mentioned in \cite{Schall:1991}, given estimate $\widehat{\xi^k}$ for $\xi^k$, we have the following updates for the maximum likelihood estimation of the variance parameters $\sigma_k^2, \; k \in \left\lbrace 1, \ldots, q \right\rbrace$:
\begin{equation*}
\sigma_{k}^2 \: \longleftarrow \:
\dfrac{\widehat{\xi^k}' A_k^{-1} \widehat{\xi^k}}
{N - \frac{1}{\sigma_{k}^2} tr\left( A_k^{-1} C_k\right)}
\quad \text{where} \quad
C_k = \left( U' W_{\xi,k} U + D_k^{-1} \right)^{-1} 
\end{equation*}

\paragraph{The algorithm}
Component $f=Xu$ is still found solving a (\ref{Jocelyn.Chauvet:CompromiseCriterion})-type program, adapting the expression of  $\psi_T$. 
Indeed, conditional on the random effects $\xi^k$, 
the working variables $z_{\xi}^k$ are assumed normally distributed according to (\ref{Jocelyn.Chauvet:working_var_assuption}). 
We thus modify the previous goodness-of-fit measure, taking into account the variance of $z_{\xi}^k$ conditional on $\xi^k$. 
In case of grouped data, 
\begin{equation}
\psi_T(u) = \sum_{k=1}^q \left\lVert z_{\xi}^k \right\rVert_{W_{\xi,k}}^2 
\cos_{W_{\xi,k}}^2 \left(z_{\xi}^k, \langle Xu, T \rangle \right)
\label{Jocelyn.Chauvet:psi_modified}
\end{equation}

\noindent
\fbox{
\begin{minipage}{\linewidth}
\begin{algorithm}[H]
\begin{description}
   \item[Step 1] Computation of the component \\
   Set: 
\begin{align*}
u^{[t]} &= 
\underset{u'Au=1}{\mbox{arg max}} \; 
\left[\phi(u)\right]^s \: \left[\psi_T(u)\right]^{1-s}  
\quad \text{where} \; \psi_T \; \text{is defined by} \; (\ref{Jocelyn.Chauvet:psi_modified}) \\
f^{[t]} &= X u^{[t]}
\end{align*}
   \item[Step 2] Henderson systems \\
For each $k \in \left\lbrace 1, \ldots, q \right\rbrace $, solve the following system:
\begin{equation*}
\begin{pmatrix}
{f^{[t]}}' W_{\xi,k}^{[t]} {f^{[t]}} & {f^{[t]}}' W_{\xi,k}^{[t]} T 
& {f^{[t]}}' W_{\xi,k}^{[t]} U \smallskip \\
T' W_{\xi,k}^{[t]} {f^{[t]}}    & T' W_{\xi,k}^{[t]} T    
& T' W_{\xi,k}^{[t]} U   \smallskip \\
U'W_{\xi,k}^{[t]}{f^{[t]}}    & U'W_{\xi,k}^{[t]}T    
& U'W_{\xi,k}^{[t]}U + {D_k^{[t]}}^{-1}
\end{pmatrix}
\begin{pmatrix}
\gamma_k \smallskip \\ 
\delta_k \smallskip \\ \xi^k
\end{pmatrix} 
= 
\begin{pmatrix}
{f^{[t]}}'W_{\xi,k}^{[t]} \, {z_{\xi}^k}^{[t]} \smallskip \\
T'W_{\xi,k}^{[t]} \, {z_{\xi}^k}^{[t]}    \smallskip \\
U'W_{\xi,k}^{[t]} \, {z_{\xi}^k}^{[t]}
\end{pmatrix}
\end{equation*}
Call $\gamma_k^{[t]}$, $\delta_k^{[t]}$ and ${\xi^k}^{[t]}$ the solutions.
   \item[Step 3] Updating variance parameters \\
For each $k \in \left\lbrace 1, \ldots, q \right\rbrace $, compute:
\begin{equation*}
{\sigma_{k}^2}^{[t+1]} =
\dfrac{ {{\xi^k}^{[t]}}' A_k^{-1} {\xi^k}^{[t]}}
{N - \frac{1}{{\sigma_{k}^2}^{[t]}} tr\left( A_k^{-1} C_k^{[t]}\right)}
\quad \text{and} \quad
D_k^{[t+1]} = {\sigma_{k}^2}^{[t+1]} A_{k}  
\end{equation*} 
   \item[Step 4] Updating working variables and weighting matrices \\
For each $k \in \left\lbrace 1, \ldots, q \right\rbrace $, compute:
\begin{align*}
{\eta^k}^{[t]} &=
f^{[t]} \gamma_k^{[t]} + T \delta_k^{[t]} + U {\xi^k}^{[t]} \\
\mu_{k,i}^{[t]} &=
G_{k}^{-1} \left({\eta_i^k}^{[t]}\right), \; i=1, \ldots, n \\
{z_{i}^{k}}^{[t+1]} &= 
{\eta^k}^{[t]} + \left(y_{i}^{k} - 
\mu_{k,i}^{[t]}\right) g_{k} \left(\mu_{k,i}^{[t]} \right), \; i=1, \ldots, n \\
W_{\xi,k}^{[t+1]} &= 
Diag\left( 
\left( 
\left[ g_{k} \left(\mu_{k,i}^{[t]} \right) \right]^{2} 
\mbox{Var} \left( Y_i^k \,|\, \xi^k \right)^{[t]}
\right)^{-1}
\right)_{i=1, \ldots, n} 
\end{align*}
\item[Step 1--4] are repeated until stability of $u$ and parameters $\gamma_k$, $\delta_k$ and $\sigma_k^2$ is reached.
\end{description}
\caption{Current iteration of the single component Mixed-SCGLR}
\label{Chauvet::PSEUDOCODE}
\end{algorithm}
\end{minipage}
}

\subsection{Extracting higher rank components}
\paragraph{}
Let $F^h = \left\lbrace f^1, \ldots, f^h \right\rbrace$ be the set of the first $h$ components.
An extra component $f^{h+1}$ must best complement the existing ones plus $T$, i.e. $T^h := F^h \cup T$.
So $f^{h+1}$ must be calculated using $T^h$ as additional covariates.
Moreover, we must impose that $f^{h+1}$ be orthogonal to $F^h$, i.e.:
\begin{center}
${F^h}' W f^{h+1} = 0$
\end{center}
Component $f^{h+1} := Xu^{h+1}$ is thus obtained solving:
\begin{equation}
\begin{cases}
\text{max} \quad  \left[\phi(u)\right]^{s} \: \left[\psi_{T^h}(u)\right]^{1-s}    \\
\text{subject to:}  \quad u' A u = 1 \; \text{and} \; {D^h}' u = 0
\end{cases}
\label{Jocelyn.Chauvet:program_plusieurs_comp}
\end{equation}
where 
$\psi_{T^h}(u) = \displaystyle{\sum_{k=1}^q} \left\lVert z_{\xi}^k \right\rVert_{W_{\xi,k}}^2 \cos_{W_{\xi,k}}^2\left( z_{\xi}^k, \langle Xu, T^h \rangle \right)$
and $D^h = X'W F^h$.

\paragraph{}
In Appendix, we give an algorithm to maximise, at least locally, any criterion on the unit sphere: the Projected Iterated Normed Gradient (PING) algorithm. 
Varying the initialisation allows us to increase confidence that the maximum reached is global.
It allows us to build components of rank $h>1$ by solving programs (\ref{Jocelyn.Chauvet:program_plusieurs_comp}) and also component of rank $h=1$ if we impose 
$T^h = T$ and 
$D^h=0$ in the aforementioned program.



\section{Simulation study in the canonical Gaussian case}
\paragraph{}
A simple simulation study is conducted to characterise the relative performances of Mixed-SCGLR and Ridge- and Lasso-based regularisations in the LMM framework 
\cite{Eliotetal:2011, GrollandTutz:2014}. We focus on the multivariate case, i.e. several $y's$, with many redundant explanatory variables.
To do so, two random responses $Y = \left\lbrace y^1,y^2 \right\rbrace$ are generated, and explanatory variables $X$ are simulated so as to contain three independent bundles of variables: $X_1$, $X_2$ and $X_3$.
Each explanatory variable is assumed normally distributed with mean $0$ and variance $1$. The level of redundancy within each bundle is tuned with parameter 
$\tau \in \left\lbrace 0.1, 0.3, 0.5, 0.7, 0.9 \right\rbrace$. 
To be precise, correlations among explanatory variables whithin bundle $X_j$ are:
\begin{equation*}
corr(X_j) = \tau \mathbf{1} \mathbf{1}' + (1-\tau) I
\end{equation*}
Besides, bundle $X_1$ ($15$ variables) models and predicts only $y^1$, bundle $X_2$ ($10$ variables) only $y^2$, while bundle $X_3$ ($5$ variables) is designed to be a bundle of noise. Considering no additional covariates ($T=0$), we thus simulate $Y$ as :
\begin{equation}
\begin{cases}
y^1 = X\beta_1 + U\xi^1 + \varepsilon^1 \\
y^2 = X\beta_2 + U\xi^2 + \varepsilon^2
\end{cases}
\label{Jocelyn.Chauvet:modele_y12}
\end{equation}
We consider the case of $N=10$ groups and $R=10$ units per group. Consequently, the random-effect design matrix can be written: $U = I_N \otimes \mathbf{1}_R$. 
All variables whithin each bundle are assumed to contribute homogeneously to predict $Y$. Then our choice of the fixed parameters are:
\begin{align*}
\beta_1 &= (\; 
\underbrace{0.3, \ldots, 0.3}_{5 \: \text{times}}, 
\underbrace{0.4, \ldots, 0.4}_{5 \: \text{times}}, 
\underbrace{0.5, \ldots, 0.5}_{5 \: \text{times}}, 
\underbrace{0, \ldots\ldots\ldots\ldots\ldots\ldots\ldots\ldots\ldots\ldots\ldots, 0}_{15 \: \text{times}}
\;)' \\
\beta_2 &=  (\; 
\underbrace{0, \ldots\ldots\ldots\ldots\ldots\ldots\ldots\ldots\ldots\ldots, 0}_{15 \: \text{times}}, 
\underbrace{0.3, .., 0.3}_{3 \: \text{times}}, 
\underbrace{0.4, .., 0.4}_{4 \: \text{times}}, 
\underbrace{0.5, .., 0.5}_{3 \: \text{times}}, 
\underbrace{0, \ldots, 0}_{5 \: \text{times}}
\;)'
\end{align*}
Finally, residual variability and within groups variability are fixed to $\sigma_k^2=1$. We thus simulate random effects and noise respectively as: 
$\xi^k \sim \mathcal{N}_N(0, \sigma_k^2 I_N)$ and 
$\varepsilon^k \sim \mathcal{N}_n(0, \sigma_k^2 I_n)$, where 
$k \in \left\lbrace 1,2 \right\rbrace$ and $n=NR$.

\paragraph{}
On the whole, $M=500$ simulations are conducted for each value of $\tau$, according to model (\ref{Jocelyn.Chauvet:modele_y12}). 
Simulation $m$ provides two fixed-effects estimations: 
$\hat{\beta}_1^{(m)}$ and $\hat{\beta}_2^{(m)}$. 
Unlike Mixed-SCGLR, both LMM-Ridge and (G)LMM-Lasso are not designed for multivariate responses: estimations are computed separately in these cases.
Consequently, for each method, we decide to retain only the one which provides the lower relative error. Their mean over $M$ simulations (MLRE) is defined as:
\begin{equation*}
\mbox{MLRE} = \dfrac{1}{M} \sum_{m=1}^M \min \left(
\dfrac{\left\lVert \hat{\beta}_1^{(m)} - \beta_1 \right\rVert^2}
{\left\lVert \beta_1 \right\rVert^2}, 
\dfrac{\left\lVert \hat{\beta}_2^{(m)} - \beta_2 \right\rVert^2}
{\left\lVert \beta_2 \right\rVert^2}
\right)
\end{equation*}

\paragraph{}
In \autoref{Jocelyn.Chauvet:res_shrinkage_par}, we summarise the optimal regularisation parameters selected via cross-validation. Corresponding MLRE's are presented in \autoref{Jocelyn.Chauvet:res_estim_study} to which we added the results provided without regularisation.
As expected, in both Ridge and Lasso regularisations, the shrinkage parameter value increases with $\tau$. On the other hand, the greater $\tau$, the more 
Mixed-SCGLR (with $l=4$ as recommended in \cite{Mortieretal:2015}) focuses on the main structures in $X$ that contribute to model $Y$.
The average value of  $s$ is approximatively $0.5$, which means that there is no significant preference between goodness-of-fit and structural relevance.
Except for $\tau=0.1$, Mixed-SCGLR provides the most precise fixed effect estimates despite the sophistication of the dependence structure and the high level of correlation among explanatory variables. 
Indeed, if there are no actual bundles in $X$ ($\tau \simeq 0$), looking for structures in $X$ may lead Mixed-SCGLR to be slightly less accurate. Conversely, the stronger the structures (high $\tau$), the more efficient our method.

\begin{table}[ht]
\begin{center}
\begin{tabular}{l p{3cm} p{3cm} p{2.75cm} p{2.5cm}}
& \centering (G)LMM-Lasso & \centering LMM-Ridge &  \multicolumn{2}{c}{\centering Mixed-SC(G)LR} \\
& \centering \small Optimal shrinkage parameter & \centering \small Optimal shrinkage parameter & \centering \small Optimal number of components $K$ & \centering \small Optimal tuning parameter $s$ \tabularnewline 
\hline \noalign{\smallskip}
$\tau=0.1$ & \centering  63.4 & \centering 24.1 & \centering 25 & \centering 0.53 \tabularnewline  
$\tau=0.3$ & \centering 111.2 & \centering 53.7 & \centering 5 & \centering 0.53 \tabularnewline 
$\tau=0.5$ & \centering 171.3 & \centering 73.2 & \centering 3 & \centering 0.51 \tabularnewline 
$\tau=0.7$ & \centering 220.6 & \centering 78.2 & \centering 2 & \centering 0.51 \tabularnewline
$\tau=0.9$ & \centering 254.9 & \centering 84.9 & \centering 2 & \centering 0.52 \tabularnewline 
\noalign{\smallskip}\hline\noalign{\smallskip}
\end{tabular}
\caption{Optimal regularisation parameter values obtained by cross-validation over $500$ simulations} 
\label{Jocelyn.Chauvet:res_shrinkage_par}
\end{center}
\end{table}

\begin{table}[ht]
\begin{center}
\begin{tabular}{l cccc} 
 & LMM & (G)LMM-Lasso & LMM-Ridge & Mixed-SC(G)LR \\ 
\hline \noalign{\smallskip}
$\tau=0.1$ & 0.141 & 0.083 & 0.090 & 0.094 \\  
$\tau=0.3$ & 0.340 & 0.180 & 0.124 & 0.105 \\ 
$\tau=0.5$ & 0.686 & 0.413 & 0.150 & 0.059 \\ 
$\tau=0.7$ & 1.571 & 0.913 & 0.189 & 0.061 \\  
$\tau=0.9$ & 5.022 & 2.431 & 0.261 & 0.050 \\ 
\noalign{\smallskip}\hline\noalign{\smallskip}
\end{tabular}
\caption{Mean Lower Relative Errors (MLRE's) associated with the optimal parameter values} 
\label{Jocelyn.Chauvet:res_estim_study}
\end{center}
\end{table}

\begin{figure}[ht]
\centering
\includegraphics[width=\linewidth]{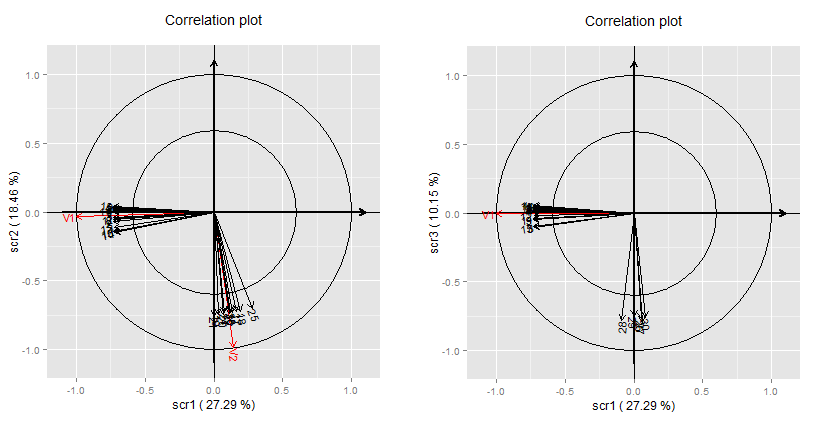}
\caption{Correlation scatterplots given by Mixed-SCGLR method on the simulated data. The left hand side and the right hand side plots respectively show component planes $(1,2)$ and $(1,3)$. Both $X$-part linear predictors related to $y^1$ and $y^2$ are considered supplementary variables.}
\label{Jocelyn.Chauvet:plans1_2_3}
\end{figure}

\paragraph{}
In order to highlight the power of Mixed-SCGLR for model interpretation, we represent on \autoref{Jocelyn.Chauvet:plans1_2_3} the correlation scatterplots obtained for 
$\tau=0.5$, $l=4$, $s=0.51$, and $K=3$. It clearly appears that $y^1$ is explained by the first bundle and $y^2$ by the second. 
The third component calculated catches the third bundle, which appears to play no explanatory role.

\section{Real data example in the canonical Poisson case}
\paragraph{}
\textit{Genus} is a dataset built from the ``CoForChange" study, which was conducted on $n=2600$ developped plots divided into $N=22$ forest concessions.
It gives the abundance of $q=94$ common tree genera in the tropical moist forest of the Congo-Basin, $p=56$ geo-referenced environmental variables, and $r=2$ other covariates which describe geology and  anthropogenic interference.
Geo-referenced environmental variables are used to represent:
\begin{itemize}
\item
$29$ physical factors linked to topography, rainfall, or soil moisture, 
\item
$25$ photosynthesis activity indicators obtained by remote sensing: EVI (Enhanced Vegetation Index), NIR (Near InfraRed channel index) and MIR (Mid-InfraRed channel index)), 
\item
$2$ indicators which describe stand of trees height.
\end{itemize}
Physical factors are many and redundant (monthly rainfalls are highly correlated, for instance, and related to the geographic location). So are all photosynthesis activity indicators. Therefore, all these variables are put in $X$. By contrast, as geology and anthropogenic interference are weakly correlated and interesting per se, we put them in the set $T$ of additional covariates.

\begin{figure}[ht]
\centering
\includegraphics[width=\linewidth]{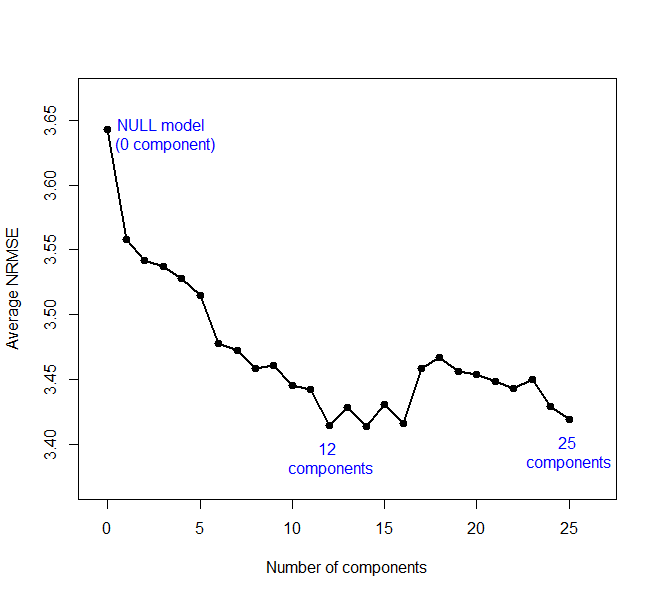}
\caption{AveNRMSE's as a function of the number of components, obtained by a cross-validation procedure. The ``null model'' does not include any explanatory variables in $X$, but only additional covariates in $T$.}
\label{Jocelyn.Chauvet:select_K_CV}
\end{figure}

\begin{figure}[ht]
\centering
\includegraphics[width=\linewidth]{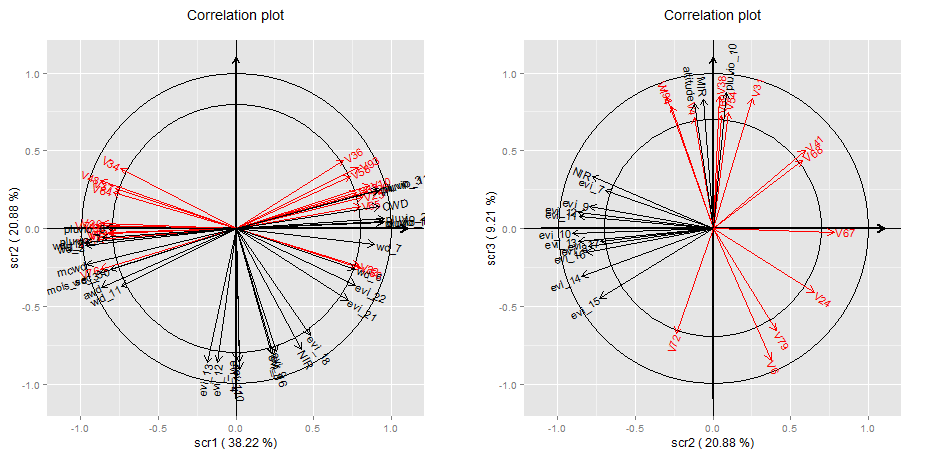}
\caption{Two examples of correlations scatterplots on data \textit{Genus}. 
The left hand side plot displays only variables having cosine greater than $0.8$ with component plane $(1,2)$. 
It reveal two patterns: a ``rain-wind''-pattern driven by the Pluvio's and Wd's variables and a photosynthesis-pattern driven by the Evi's.
On the right hand side, we plot variables having cosine greater than $0.7$ with component plane $(2,3)$. 
Component $3$ reveals a bundle driven by variables Altitude, MIR and Pluvio10, which prove important to model and predict several $y$'s.}
\label{Jocelyn.Chauvet:two_first_circles}
\end{figure}

\paragraph{}
It must be noted that the abundances of species given in \textit{Genus} are count data. For each random response $y^1, \ldots, y^q$ we thus choose a Poisson regression with $\log$ link: 
\begin{equation*}
\forall k \in \left\lbrace 1, \ldots, q \right\rbrace, \quad
y^k \sim \mathcal{P} \left( 
\exp \left[
\sum_{j=1}^K \left( Xu^j \right) \gamma_j + T\delta_k + U\xi^k
\right]
\right)
\end{equation*}
Among a series of parameter choices, values $l=4$ and $s=0.5$ prove to yield components very close to interpretable variable bundles.
We therefore keep these parameter values in order to find - through a cross-validation procedure - the number of components $K$ which minimises the Average Normalised Root Mean Square Error (AveNRMSE) defined as:
\begin{equation*}
\mbox{AveNRMSE} = \dfrac{1}{q} \sum_{k=1}^q
\sqrt{\dfrac{1}{n} \sum_{i=1}^n \left( \dfrac{y_i^k - \hat{y}_i^k}{\bar{y}^k} \right)^2}
\end{equation*}
On \autoref{Jocelyn.Chauvet:select_K_CV}, we plot the AveNRMSE's for $K \in \left\lbrace 0, 1, \ldots, 25 \right\rbrace$. As one can see, the best models are the ones with $12$, $14$ and $16$ components. 
We retain the most parcimonious of them, i.e the one with $12$ components. Two examples of correlation scatterplots we obtain are given on \autoref{Jocelyn.Chauvet:two_first_circles}, in which the $X$-parts of linear predictors are considered supplementary variables. 

\section{Conclusion}
Mixed-SCGLR is a powerful trade-off between multivariate GLMM estimation (which cannot afford many and redundant explanatory variables) and PCA-like methods (which take no account of responses in building components).
While preserving the qualities of the plain version of SCGLR, the mixed one 
performs well on grouped data, and 
provides robust predictive models based on interpretable components.
Compared to penalty-based approaches as Ridge or Lasso, the
orthogonal components built by Mixed-SCGLR
reveal the multidimensional explanatory and predictive structures, and greatly facilitate the interpretation of the model.

\section*{Acknowledgement}
The extended data \textit{Genus} required the arrangement and the inventory of 140.000 developped plots across four countries : Central African Republic, Gabon, Cameroon and Democratic Republic of Congo. The authors thank the members of the CoForTips project for the use of this data.

\section*{Appendix}

The Projected Iterated Normed Gradient (PING) is an extension of the iterated power algorithm, solving any program which has the form:
\begin{equation}
\max_{\underset{D'u=0}{u'Au=1}} h(u)
\label{Jocelyn.Chauvet:genericPING1}
\end{equation}
Note that putting $v := A^{1/2}u$, $g(v) := h(A^{-1/2}v)$ and $C := A^{-1/2}D$, program (\ref{Jocelyn.Chauvet:genericPING1}) is strictly equivalent to program (\ref{Jocelyn.Chauvet:genericPING2}):
\begin{equation}
\max_{\underset{C'v=0}{v'v=1}} g(v)
\label{Jocelyn.Chauvet:genericPING2}
\end{equation}
In our framework, the particular case $C=0$ (no extra-orthogonality contrain) allows us to find the first rank component.
Denoting $\Pi_{C^{\bot}} := I - (C'C)^{-1} C'$ and  
$\Gamma \left( v \right) := \underset{v}{\nabla} \: g(v)$, 
a Lagrange multiplier- based reasoning gives the basic iteration of the PING algorithm:
\begin{equation}
v^{[t+1]} = \dfrac{\Pi_{C^{\bot}} \Gamma \left( v^{[t]} \right)}
{\left\lVert \Pi_{C^{\bot}} \Gamma \left( v^{[t]} \right) \right\rVert}
\label{Jocelyn.Chauvet:basinPINGiter}
\end{equation}
Despite the fact that iteration (\ref{Jocelyn.Chauvet:basinPINGiter}) follows a direction of ascent, it does not guarantee that $g$ actually increases on every step.
Algorithm PING therefore repeats the following steps until convergence of $v$ is reached:
\begin{description}
   \item[Step 1] Set: 
   $\kappa^{[t]} = \dfrac{\Pi_{C^{\bot}} \Gamma \left( v^{[t]} \right)}
{\left\lVert \Pi_{C^{\bot}} \Gamma \left( v^{[t]} \right) \right\rVert}$
   \item[Step 2] A Newton-Raphson unidimensional maximisation procedure is used to find the maximum of $g(v)$ on the arc $\left( v^{[t]}, \kappa^{[t]} \right)$ and take it as $v^{[t+1]}$.
\end{description}


\begin{thebibliography}{99}

\bibitem{Batienetal:2004} Bastien, P., Esposito Vinzi, V. and Tenenhaus, M. (2004)
\textit{PLS generalized linear regression.} Computational Statistics \& Data Analysis, \textbf{48}, 17--46.

\bibitem{Bryetal:2013} Bry, X., Trottier, C., Verron, T. and Mortier, F. (2013) 
\textit{Supervised component generalized linear regression using a PLS-extension of the Fisher scoring algorithm.} Journal of Multivariate Analysis, \textbf{119}, 47-–60.

\bibitem{Eliotetal:2011} Eliot, M., Ferguson, J., Reilly, M.P. and Foulkes, A.S. (2011) 
\textit{Ridge Regression for Longitudinal Biomarker Data.} The International Journal of Biostatistics, \textbf{7}, 1--11.

\bibitem{GrollandTutz:2014} Groll, A. and Tutz, G. (2014) 
\textit{Variable Selection for Generalized Linear Mixed Models by L1-Penalized Estimation.} Statistics and Computing, \textbf{24}, 137--154. 

\bibitem{Henderson:1975} Henderson, C.R. (1975)
\textit{Best linear unbiaised estimators and prediction under a selection model.} Biometrics, \textbf{31}, 423–-447. 

\bibitem{Marx:1996} Marx, B.D. (1996) 
\textit{Iteratively reweighted partial least squares estimation for generalized linear regression.} Technometrics, \textbf{38}, 374–-381.

\bibitem{McCullochSearle:2001} McCulloch, C.E. and Searle, S.R (2001) 
\textit{Generalized, Linear, and Mixed Models.} John Wiley \& Sons.

\bibitem{Mortieretal:2015} Mortier, F., Trottier, C., Cornu, G., Bry, X. (2015)
\textit{SCGLR - An R Package for Supervised Component Generalized Linear Regression.} 

\bibitem{Schall:1991} Schall, R. (1991)
\textit{Estimation in generalized linear models with random effects.} Biometrika, \textbf{78}, 719–-727.

\bibitem{Stiratellietal:1984} Stiratelli, R., Laird, N., and Ware, J.H. (1984) 
\textit{Random-Effects Models for Serial Observations with Binary Response.} Biometrics, \textbf{40}, 961--971.

\end{thebibliography}
\end{document}